\documentclass[11pt]{article}
\usepackage{amsmath}
\usepackage[margin=1in]{geometry}
\usepackage{color}
\usepackage{graphicx,subfig}
\usepackage{authblk}

\newcommand*\samethanks[1][\value{footnote}]{\footnotemark[#1]}

\title{\LARGE{\textbf{DRAFT}}\\\vphantom{}Heat conduction through permafrost and its potential for explosive behavior}
\author{Kaitlin Hill\thanks{Email: hillk@umn.edu} \thanks{School of Mathematics, University of Minnesota, Minneapolis, MN 55455} }
\author{Richard McGehee\samethanks[2]}
\affil{}
\date{}

\begin{document}

\maketitle

\begin{abstract}
\noindent The recent widespread thaw of permafrost has led to observations of explosive gas emissions, which expel ice and soil debris and leave behind large craters. This phenomenon appears to be caused by a buildup of pressure from below the permafrost, possibly due to gas released as permafrost melts, followed by a sudden emission of gas through the surface. Although there have been some studies modeling the processes involved in crater formation using computationally complex models, we propose that these explosive events can be attributed to a simple heat diffusion-based process. Under certain boundary conditions and parameters, this may be sufficient to describe the explosive behavior observed. We demonstrate this effect by linearly increasing surface temperature from average monthly values (1961-1990) at an example latitude, which causes more dramatic melting from below the permafrost than above. This may lead to a buildup of gas pressure, if the permafrost is both continuous and has a high ice saturation, and has the potential for sudden gas release.
\end{abstract}

\section{Background}

Along with permafrost melt in the past decade, there have been observations of a previously-unseen phenomenon, currently described as a `permafrost crater'. Permafrost is defined as ground where temperatures have remained below 2$^{\circ}$C for at least two consecutive years. Permafrost craters, which have been primarily observed in Siberia \cite{Buldovicz, Kizyakov-Comparison, Kizyakov-Microrelief, Leibman}, are believed to be created by explosive gas emissions from melting permafrost. Observations of explosive gas emissions have also been studied on the Arctic sea floor and anecdotally mentioned in studies of sub-permafrost gas collection in Northwest China \cite{Andreassen, Wang-Qilian}. Observations of crater or funnel formation are currently limited to remote sensing from satellite and post-formation reconstructions \cite{Kizyakov-Comparison, Kizyakov-Microrelief, Leibman}.  

Permafrost craters can range from $25-100$ m wide, with a parapet of soil piled up to 4 m high around the edge of the crater. The depth of the craters may vary, and many become filled with water over a period of a few years. Several observational studies note that they occur in locations where permafrost is continuous and the active layer is about 1 m deep, and that these explosions eject blocks of ice and permafrost \cite{Leibman}.

There have been few mathematical studies of potential influences on crater formation, and those that exist are primarily empirically-derived \cite{Buldovicz, Khimenkov, Wang-Qilian}. Although models exist for the diffusion of methane through permafrost, they do not explore the potential for explosive behavior \cite{Kaiser, Nakano}.


In this note we propose a simple heat conduction-based model for permafrost melt through a column of lithosphere. 
This model provides a highly idealized description of permafrost melt compared to current studies of heat conduction through the lithosphere \cite{Riseborough}.  However, we contend that the possibility of sudden gas release is present even in this toy model, which indicates that simple processes may govern this sudden release. Experimental evidence suggests that a build-up of pressure may be a primary factor in the explosive behavior causing permafrost craters \cite{Yakushev}. 
Using this conceptual model, we hypothesize that permafrost melt occurs both in the active layer and at the lower permafrost boundary, and that the melting at the lower boundary may cause a build-up of pressure due to gas release below the permafrost, if the permafrost has a high ice saturation. The combination of a thin layer of permafrost and pressure from methane gas release below this layer may be sufficient to cause an explosive emission of methane at the surface.

\section{A simple model of heat diffusion}
We consider a simple column model for heat conduction through the mantle in the presence of permafrost. For a given latitude, we model the vertical diffusion of heat through the soil as a heat equation with Dirichlet boundary conditions,
\begin{align*}
	\frac{\partial T}{\partial t} &= k \frac{\partial^2 T}{\partial z^2}, \qquad ,\quad t>0\quad ,\quad 0<z<l \\
	T(z,0) &= \dfrac{M-T_S(0)}{l}z + T_S(0), \\
	T(0,t) &= T_S(t) + F, \\
	T(l,t) &= M, 
\end{align*}
where $k$ is the thermal diffusivity. Here, we have assumed heat travels vertically in a column through the earth, with $z=0$ at the surface and increasing downward. We initialize the system using the equilibrium temperature profile of the model with no added temperature forcing ($F\equiv 0$).

\begin{figure}[t!]
	\centering
	\subfloat[]{\includegraphics[scale=0.9]{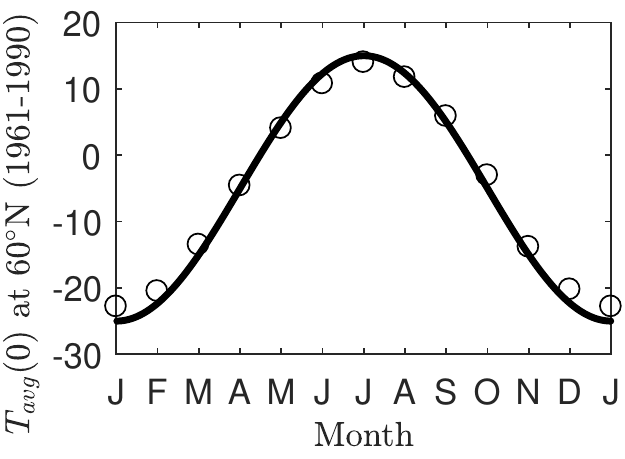}}
	\qquad\subfloat[]{\includegraphics[scale=0.9]{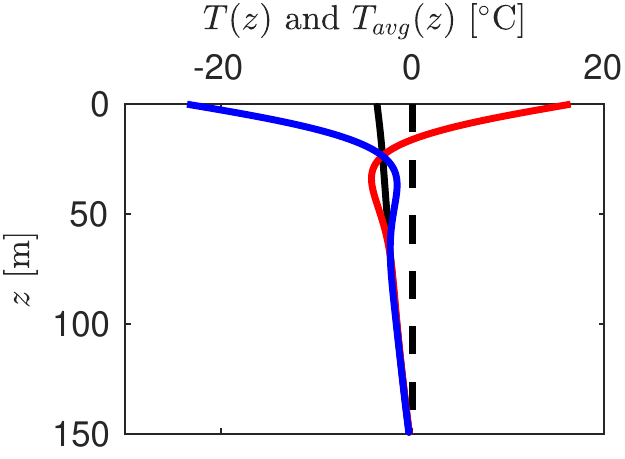}} \\
	\subfloat[]{\includegraphics[scale=0.9]{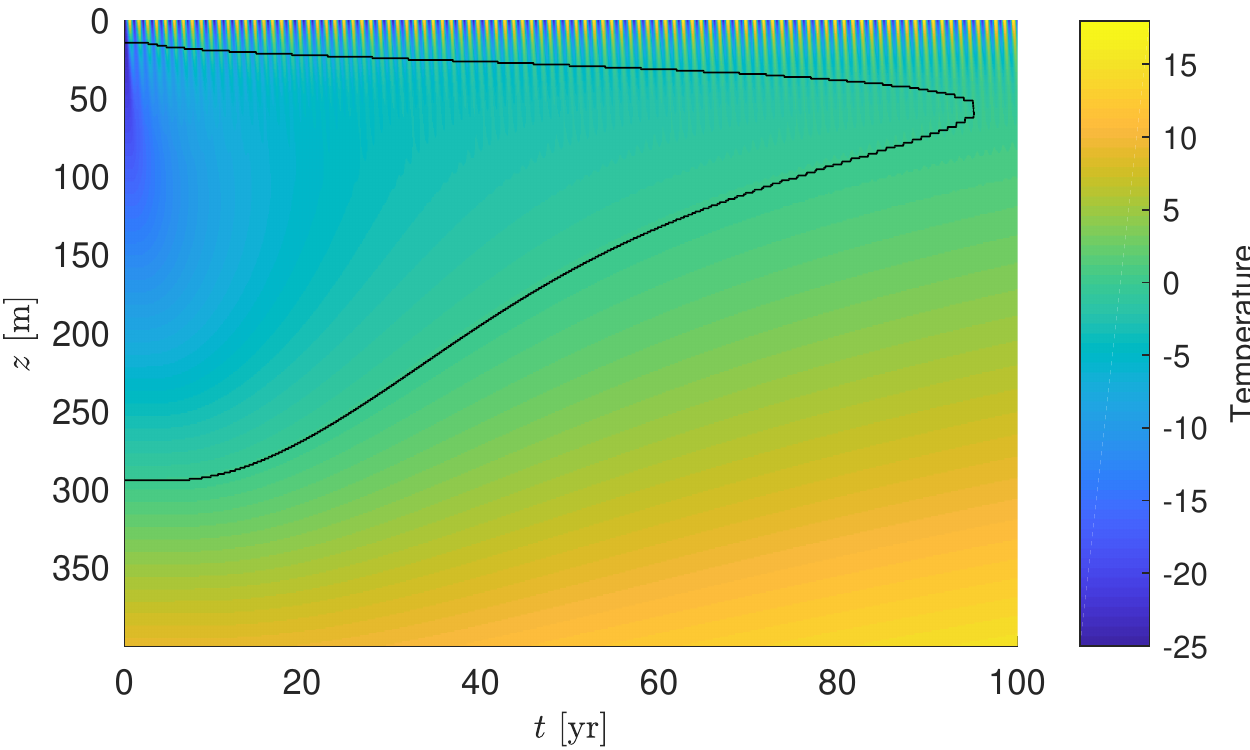}}
	\caption{(a) Surface temperature $T_S(t)$ approximation (curve) and monthly average temperature observations at $61^{\circ}$N (scatter points). Observations calculated from land surface temperatures in CRU CL v2.0 \cite{New}. (b) Profile of the maximum (red), minimum (blue), and average (black) temperature during the year beginning at $t=50$ years. (c) Permafrost depth variation as $F$ increases, where the color gradient shows the temperature evolution and the black line gives the permafrost boundary. Simulations use the parameters $k=700,$ $M=60,$ and $l=1,000$.}
	\label{fig:1}
\end{figure}

The boundary condition at the surface, $T(0,t)=T_S(t)+F,$ represents a combination of the temperature at the surface $T_S(t)$ and a constant forcing that represents a linear increase in temperature. This temperature increase may be attributed in part to rising greenhouse gases. In general, increasing greenhouse gases causes an approximately logarithmic reduction in outgoing longwave radiation at the surface, leading to a corresponding increase in temperature \cite{Pierrehumbert}. 
Although in principle the model applies to any surface temperature function, we approximate the temperature at the surface as sinusoidal with period one year, $T_S(t)\equiv -5-20\cos(2\pi t)$. Figure~\ref{fig:1}(a) compares our approximation to monthly mean land surface temperature observations, averaged from 1961 to 1990 \cite{New}. We use surface temperature averages at approximately 61$^{\circ}$N, as an example latitude near the edge of continuous permafrost.

We take the boundary condition from below to be a heat source from the convective portion of the mantle. We approximate the temperature at the lower boundary of the crust as constant, $M\in [10,35]$ $^{\circ}$C at a depth of $l=1,000$ m \cite{Earle,Slagstad}. For the purposes of simulating a thin layer of permafrost, however, we set $M=60$. Although this is higher than the expected temperature at 1,000 m depth, this provides an initial relatively thin layer of permafrost without incorporating additional complexity into the model.

%

The thermal diffusivity coefficient is defined as $k\equiv K/\rho c,$ where $\rho c \approx 0.5$~cal~cm$^{-3}$~K$^{-1}$ is the volumetric heat capacity of the medium and $K$ is its thermal conductivity \cite{Lachenbruch}. Volumetric heat capacity is approximately 0.5~cal cm$^{-3}$~K$^{-1}$~$\approx 0.06$~W~yr~m$^{-3}$~K$^{-1}$ for all soils including ice \cite{Lachenbruch}. The thermal conductivity of soil near freezing varies dramatically depending on the moisture content and dry density of the soil \cite{Farouki, Johansen, Kersten}. Following the meta-analysis of Farouki, we let $K\in[5,55]$~W~m$^{-1}$~K$^{-1}$, so $k\in [75, 828]$~m$^2$~yr$^{-1}$ \cite{Farouki}. The lower end of the range corresponds to 0.25 degrees of saturation, a temperature of $-4^{\circ}$C, and the thermal conductivity of the solid part of the soil is 2 W m$^{-1}$ K$^{-1}$. The upper end of the range corresponds to fully saturated soil (1.0 degree of saturation), a temperature of $-4^{\circ}$C, and a thermal conductivity of the solid part of the soil of 9~W~m$^{-1}$~K$^{-1}$.


\section{Potential for explosion}

In this section we give a brief overview of model behavior as temperature is increased linearly by increasing $F$. We give a qualitative evaluation of the model spatial profile and temperature variation through a column of soil. 

We simulate the effect of increased greenhouse gases by ramping the parameter $F$ linearly from $0^{\circ}$C to 3$^{\circ}$C over a period of 100 years. Figure \ref{fig:1}(b) shows the temperature variation over a typical year mid-way through the simulation, where there is still an active layer in the permafrost. While the model does not involve enough physics to give a quantitatively accurate temperature profile, it is able to return one that is a qualitatively similar.

Figure \ref{fig:1}(c) shows the evolution of the temperature profile over the entire simulation, with the permafrost boundaries indicated by the black curve. As depth increases from the surface, the permafrost begins above a thin permafrost-free active layer, and ends at a lower boundary formed by heat flux coming from the mantle. Note that although increasing $F$ changes the boundary condition at the surface, the permafrost is first appreciably affected at the lower boundary of the permafrost.

The permeability of methane through permafrost becomes minimal (1/100 darcys) at 60\% saturation and negligible at 80\% saturation; see \cite{Wang-Qilian}. Given a permafrost layer that is initially thin, the combined effect of an initial imbalance of greater warming from below than above, along with highly saturated permafrost, may lead to a build-up of pressure due to gas release from beneath the permafrost. When this pressure reaches a critical level, it may release suddenly and explosively through a weak point in the permafrost layer.

\end{document}